# Self-Assembling Ice Membranes on Europa: Brinicle Properties, Field Examples, and Possible Energetic Systems in Icy Ocean Worlds

Steven D. Vance,[1] Laura M. Barge,[1] Silvana S.S. Cardoso,[2] and Julyan H.E. Cartwright[3,4]

[1]NASA Jet Propulsion Laboratory, California Institute of Technology, Pasadena, California, USA.

[2]Department of Chemical Engineering and Biotechnology, University of Cambridge, Cambridge, UK.

[3]Instituto Andaluz de Ciencias de la Tierra, CSIC–Universidad de Granada, Granada, Spain.

[4]Instituto Carlos I de Física Teórica y Computacional, Universidad de Granada, Granada, Spain.

**Abstract**

Brinicles are self-assembling tubular ice membrane structures, centimeters to meters in length, found beneath sea ice in the polar regions of Earth. We discuss how the properties of brinicles make them of possible importance for chemistry in cold environments—including that of life's emergence—and we consider their formation in icy ocean worlds. We argue that the non-ice composition of the ice on Europa and Enceladus will vary spatially due to thermodynamic and mechanical properties that serve to separate and fractionate brines and solid materials. The specifics of the composition and dynamics of both the ice and the ocean in these worlds remain poorly constrained. We demonstrate through calculations using FREZCHEM that sulfate likely fractionates out of accreting ice in Europa and Enceladus, and thus that an exogenous origin of sulfate observed on Europa's surface need not preclude additional endogenous sulfate in

Europa's ocean. We suggest that, like hydrothermal vents on Earth, brinicles in icy ocean worlds constitute ideal places where ecosystems of organisms might be found. Key Words: Europa—Brinicles—Icy ocean worlds—Ice membranes—Chemical gardens. Astrobiology 19, xxx–xxx.

## 1. Introduction

PHYSICAL AND CHEMICAL GRADIENTS are essential for life: they provide the setting where life can thrive between redox potentials, using this ambient energy to promote biosynthesis. Certain communities of organisms known today feed directly off environmental disequilibria such as the gradients of redox substrates that exist at sediment-water interfaces on Earth. Life in such systems *produces* chemical and electrochemical gradients in the environment, which can themselves be biosignatures (Shock and Canovas, 2010). The ocean-seafloor interface is the most well-known example (*e.g.,* Kelley *et al.,* 2007), but the interface between ice and seawater may also facilitate life (Trinks *et al.,* 2005; Attwater *et al.,* 2010; Krembs *et al.,* 2011; Bartels-Rausch *et al.,* 2012). In ice-covered ocean worlds such as Jupiter's moon Europa, the ice-ocean interface may be able to sustain the highest level of biological productivity if oxidizing fluids generated in the ice flow into a reducing ocean (Russell *et al.* 2017).

Beyond Earth, physicochemical interfaces may have unfamiliar compositions and spatial scales that require generalized models inspired by potentially analogous habitats on Earth. For example, in the ocean on Jupiter's moon Europa, oxidants percolating, diffusing, or advecting through the global ice covering may be the best source of energy for life when combined with reductants produced at the seafloor (Hand *et al.,* 2007, 2009; Vance *et al.,* 2016). The temperatures of brines flowing through ice may reach the eutectic values for sulfuric acid (211 K; McCarthy *et al.,* 2007) or ammonia (175 K; Kargel, 1992; Choukroun and Grasset, 2010), while temperatures in supercritical fluids emerging from hydrothermal edifices at the seafloor may exceed the typical values of 350 K obtained in systems studied on Earth (Lowell and Dubose, 2005). Studies of the properties of generalized environmental interfaces, including their formation processes and energetic characteristics, can aid in understanding not only life on Earth but also where best to search for life on other worlds.

Chemical-garden structures provide a general model for physicochemical interfaces that may support life. They form at the interfaces between contrasting fluids at hydrothermal vents and are easily created in the laboratory. These chemobrionic structures self-assemble in diverse conditions of buoyancy, internal fluid pressure (*e.g.,* hydrothermal injection rate), osmotic force, and gradients in composition between the ocean and hydrothermal fluids (Barge *et al.,* 2015a). Natural chemobrionic membranes (Barge *et al.,* 2015a) are porous, semipermeable (Filtness *et al.,* 2003; Mielke *et al.,* 2011), electrically conductive and catalytic (Nakamura *et al.,* 2010; Yamamoto *et al.,* 2013). They can sequester other reactive ions into the solid phase (Barge *et al.,* 2015b) and can generate membrane potentials from selective permeability and the precipitation of ions (Barge *et al.,* 2012, 2015c; Glaab *et al.,* 2012). In some ways, chemobrionic membranes such as those in hydrothermal chimneys have similar functionality to biological membranes: they maintain pH and redox potentials between two reservoirs and thus maintain disequilibria in a system that would otherwise dissipate its energy through mixing and homogenization. Although chemical-garden systems are not alive, they behave similarly in some ways to biological systems (Leduc, 1911), and in geological settings they can mediate energy and host ecosystems. As such, they are hypothesized to be significant in the origin of life on Earth and possibly on other worlds (Russell and Hall, 1997; Martin and Russell, 2007; Russell *et al.,* 2010, 2014, 2017).

Another possible fount for life's origin is the icy analogs of chemical gardens, or brinicles (Dayton and Martin, 1971; Eide and Martin, 1975; Cartwright *et al.,* 2013), a less studied geological example of physicochemical interfaces. Also known as ice stalactites, they are tubes of ice formed from dense supercooled brine flowing downward out of ice sheets (Fig. 1). Brinicles have been observed during the winter in Antarctica, beneath the pack ice and close enough to land that strong ocean currents do not break them (Paige, 1970). They have also been found in similar environments in the northern hemisphere (Sakshaug *et al.,* 1994; Trinks, 2001).

Their properties have been studied experimentally (Martin, 1974). Salt separates out of the ice under the force of gravity, traveling through osmosis and along established brine channels (Lake and Lewis, 1970; Fig. 2). The downflowing brine absorbs heat from the surrounding ocean and freezes the seawater around itself, forming a tube. The tube continues to grow while brine flow extracts enough heat from the surrounding water to freeze more ice. Figure 3 illustrates the resulting structure, with comparisons for Europa and Enceladus based on the calculations described below. Like chemical gardens, the membrane ice separates contrasting chemical environments: the interior being a brine with very high dissolved ion content and possibly a different pH than the exterior seawater. The formation of ice creates a gradient in the activity of water near the solid-liquid interface. Also, ice membranes may be capable of generating ion potentials between the two fluids, due to ice's ability to preferentially conduct protons (Petrenko and Whitworth, 1999). Ice membranes, like chemical gardens, are semipermeable, creating osmotic forces (Miller, 1970). Thus, we argue that brinicle formation should be studied as a self-assembly process analogous to chemical gardens in geological systems, and the properties of these ice membranes should be characterized to understand whether they can function as energy-generating systems for life.

In this paper, we focus our discussion on the physical, chemical, and electrochemical properties of ice chemical gardens and compare them with other more well-understood self-assembling chemical systems. We discuss how ice membranes could act as energy generators and habitats for life on Earth and, possibly, on other worlds such as Europa.

## 2. Properties of Ice Membranes

The formation of ice-membrane structures has some interesting energetic implications, judging by the analogous processes in chemical precipitation systems that generate pH gradients and ion potentials. In a chemical-garden structure, the precipitation of insoluble components at the fluid interface produces a Nernst potential between the two solutions that is determined by the remaining concentration gradients of ions, including $OH^-$ and $H_3O^+$, across the membrane (Barge *et al.*, 2015a; Cartwright *et al.*, 2002). This can be measured as a membrane potential. Experimental studies show that in metal-silicate systems membrane potentials can approach 100–200 mV (Barge *et al.*, 2012; Glaab *et al.*, 2012), and in iron-sulfide systems they can reach 500–600 mV (Russell and Hall, 2006; Barge *et al.*, 2015a, 2015c; Barge and White, 2017). Membrane potentials may also be presumed to occur in geological chemical-garden precipitates like the metal-sulfide chimneys formed at black smoker vents (Yamamoto *et al.*, 2013, 2017; Barge *et al.*, 2015b).

Ice membranes may generate useful potentials in a similar way. Brinicle structures, though not formed by ion precipitation, still separate solutions of very different salinity and so should create gradients of $Na^+$, $Cl^-$, $Ca^{2+}$, $Mg^{2+}$, $SO_4^{2-}$, and other dissolved salts across the membrane wall. Gradients in pH might also develop as the brines produced in the ice shelf change pH as they become concentrated relative to seawater (Trinks, 2005). Chemical-garden membranes generate potential differences by being more permeable to some ions than others, and ice membranes might generate potential differences by functioning as proton exchange membranes. It is known that $H_3O^+$ permeates for a greater distance into ice than $OH^-$ (Dash and Wettlaufer, 2003; Moon *et al.*, 2012), so pH gradients might be spontaneously generated and maintained across brinicle membranes. Ice membranes are also capable of generating osmotic forces, as shown in an osmotic filtration experiment by Miller (1970). In this experiment, an ice membrane was clamped between two solutions of different compositions, and reverse osmosis was performed by imposing an external pressure. Such osmotic behavior enables the brinicle, like the chemical garden, to be powered by an osmotic pump (Cartwright *et al.*, 2013).

In geological chemical-garden systems, there is also the issue of possible electron transfer if the membrane wall is composed of conductive material and there is an ambient electronic potential between the exterior and interior solutions. This can occur, for example, when reduced compounds produced by hydrothermal synthesis interface with seawater $O_2$ (Nakamura *et al.*,

2010; Yamamoto *et al.,* 2013). Ice is not very electrically conductive unless it contains impurities such as ionic inclusions (Gross *et al.,* 1978), and the ice composing a self-assembling tubular structure would likely exclude salt ions. If an ice membrane separated two solutions containing compounds of different redox potential, there would likely be little to no electron transfer and redox catalysis driven by the ice membrane itself, as can occur in the walls of hydrothermal chimney precipitates. However, while ion and electron transport are inhibited, ice is a proton conductor, translocating protons through the crystal lattice via the Grotthuss mechanism (de Grotthuss, 1806; Cukierman, 2006). The walls of a brinicle might therefore function as a proton exchange membrane, allowing proton ($H_3O^+$) diffusion from one solution to another but maintaining the other chemical and redox gradients across the wall. The surfaces of the ice membrane might in this way be able to drive biochemistry by donating or accepting protons, or producing a proton-motive force at certain locations if the chemical conditions were right.

## 3. Brinicle Structures as Habitats

Chemical-garden structures are a general phenomenon emerging from fluid interfaces in far-from-equilibrium chemical systems, and they can occur at many different scales (Barge *et al.,* 2015a). Experimentally they can be formed in a test tube, but in natural systems they can be anywhere from submicrometer-sized (*e.g.,* cement tubes; Cardoso *et al.,* 2017) to tens of meters tall with decimeter internal diameter (*e.g.,* hydrothermal chimneys at alkaline vents or black smokers [Ludwig *et al.,* 2006; Kelley *et al.,* 2007; Ding *et al.,* 2016]). The scale of the tubular structure that forms is related to the physical and chemical parameters of the system, such as reactant concentrations, fluid flow and internal pressure, precipitation conditions and solubility of membrane material (Kaminker *et al.,* 2014). The formation of the precipitation structure guides the further flow of fluid, and if the precipitate membrane can isolate the inner fluid from the exterior reactants for some distance, then the inner solution will continue to flow through the structure until it can emerge at the top and induce precipitation of dissolved ions. Thus, the huge chemical gardens that grow at hydrothermal vents—such as the 60 m tall (but with micrometer-scale tubes) carbonate chimneys at the Lost City hydrothermal vent field (Kelley *et al.,* 2005; Ding *et al.,* 2016)—are a product of tens of thousands of years of a chemical gradient being maintained through geological processes (Früh-Green *et al.,* 2003) and can serve as habitats and gradient-mediating structures for a very long time.

Brinicles discovered so far in Antarctica are meter-scale in length and 10 or more centimeters in external diameter. They are more fragile than other chimneys: they are dislocated by strong ocean currents and thus can only form below the sea ice closer to land (Paige, 1970). They are also not long-lived, since they depend on the winter air temperature to produce the supercooled brine that feeds growth of the tube, so they disappear with seasonal changes (Dayton and Martin, 1971; Cartwright *et al.,* 2013). However, as with the mineral precipitates at hydrothermal vents, it is possible to imagine situations in which tubular ice structures growing downward from an ice sheet grow thicker and thus more stable. The ice membrane wall grows via heat loss from seawater directly interfacing the brine stream, so the temperature gradient between the seawater and brine, along with the thermal conductivity of the ice, determines how thick the wall will become and correspondingly how far the brinicle will grow. Heat from the interior brine fluid must either be dissipated through the ice wall (thus freezing additional seawater and making the wall grow thicker) or dissipated as the brine reaches the outlet of the tube, thus lengthening the whole tube from that outlet point. In Antarctica, the tubes can reach all the way from the ice sheet to the seafloor (Jeffs, 2011).

Tubular structures can serve as channels for nutrient delivery. For example, in serpentinizing hydrothermal vent systems, $H_2$ is produced and delivered as a fuel for a whole ecosystem that lives in that vent (Lang *et al.,* 2012). Such systems may persist for tens of thousands of years (Lowell and Rona, 2002; Früh-Green *et al.,* 2003). The tubes constitute a "funnel" that concentrates the substrates/ions; and in a brinicle the tip of the brinicle would deliver a large concentration of salt ions and macronutrients compared to the surrounding seawater. Redox reactants, which in another mineral precipitate might transfer electrons directly through the wall, as we considered above, would in an ice structure flow all the way to the end of the tube. This would provide a microenvironment at the outlet point where oxidants (from the brine, for example) and reductants (from the ocean, for example) would both be provided and concentrated, within easy reach for an organism (Shock and Canovas, 2010).

## 4. Brinicles in Icy Ocean Worlds

Here we discuss the physical magnitude of liquid water expulsions in icy worlds, and the possible occurrence of brinicle structures that persist over geologically significant time scales. In sea ice on Earth, brine expulsions occur on a seasonal basis as insolation decreases with the onset of polar winter. Multiple meters of ice are generated in a matter of weeks (Bartels-Rausch *et al.,* 2012), containing brine pockets with well-defined concentrations determined by the thermodynamics of freezing in aqueous solutions (Marion *et al.,* 1999; Butler and Kennedy 2015). In extraterrestrial oceans, brine expulsions at the ice-ocean interface should occur in proportion to the rate of overturning of the ice, and possibly also on the diurnal timescale of the tides. Here we consider the example of Europa.

Europa's ice may drive the formation of brinicles as a delivery mechanism for materials delivered to the surface from space. The ice appears to contain many irradiated impurities (Carlson *et al.,* 2009), mainly from neighboring Io, which should cycle through the ice on a time scale shorter than the estimated mean surface age of 60–100 Myr (Zahnle *et al.,* 2008). Melt should occur within the ice, as present-day tidal heating in the ice likely exceeds dissipation in the mantle and the adiabatic temperature gradient within the ice should be close to the melting curve (Tobie *et al.,* 2003; Sotin *et al.,* 2009). The attenuation at small tidal frequencies appears to be less grain-size dependent than has been often assumed, which may lead to melting more readily than previously supposed (McCarthy and Cooper, 2016). Upwelling warm ice has been linked to chaotic terrain at Europa's surface, at least one site of which may have been active when *Galileo* last imaged them (Schmidt *et al.,* 2011). Fluids may also occur along linear faults in Europa's ice owing to tidal flexing (Kalousová *et al.,* 2016). Depending on the temperature and porosity profile within the ice, dense near-surface fluids may transfer to the ocean on the $10^5$ yr timescale of solid-state convection (Sotin *et al.,* 2002; Nimmo and Giese, 2005; Kalousová *et al.,* 2014).

Separate from transport from the surface, the ice-ocean interface may concentrate brines within the lower regions of the ice due to regional variations in local heating from below. This heat may come from hydrothermal activity at the seafloor (Goodman *et al.,* 2004; Vance and Goodman, 2009; Goodman and Lenferink, 2012) or from global Hadley circulation (Soderlund *et al.,* 2014). Large-scale variations in ice thickness may lead to latitudinal overturning circulation and saline stratification (Ashkenazy *et al.,* 2017; Zhu *et al.,* 2017). The ice shell thickness has been mapped at Enceladus and ranges from 5 to 40 km (Čadek *et al.,* 2016). On Europa, the ice thickness is expected to vary laterally by no more than 7 km (Nimmo *et al.,* 2007). On a timescale of tens of hours, the equilibrium freezing pressure near the ice-ocean interface should move with the tidal bulge, which on Europa has a height around 30 m (Moore and Schubert, 2000). These processes imply melting and reaccretion of fresh or so-called marine ice.

The ionic content of the ice should vary with depth due to equilibrium and kinetic fractionation, and due to differential transport through diffusion. This underscores the importance of understanding these processes when seeking to infer the ocean's as-yet-unknown salinity (both composition and concentration; Carlson et al., 2009; Zolotov and Kargel, 2009). Where accretion occurs under the ice, oceanic brines should concentrate out near the ice-ocean interface (Marion *et al.,* 2005; hereafter M2005). For a thermally conductive vertical profile in the ice, brines should freeze out within the lower 10% of the ice thickness from the ice-ocean interface because the remaining 90% of the overlying ice will be below the eutectic temperature. For convective ice, the eutectic horizon will be even closer to the ice-ocean interface if the characteristic temperature of the convecting region, $T_c$, is less than the eutectic temperature, and very close to the surface for brine mixtures with eutectic temperatures above $T_c$.

In Fig. 4, we reproduce the M2005 Europa equilibrium freezing calculation for a conductively cooling 20 km thick ice shell (Fig. 4 from that work) using FREZCHEM v15.1. The initial and final mass of water is fixed at 1 kg, as per M2005. Also as per M2005, we fix the temperature at the bottom of the ice to 263 K, so our calculations do not display intermediate fractionation for less concentrated solutions that initially increases the concentration of all brines within the ice. Whereas M2005 considered only the Mg and $SO_4$ dominated briny ocean from the work of Kargel *et al.* (2000; Kargel2000), here we also investigate the bulk silicate Earth (BSE) K1 and K2 models for Europa from Zolotov and Shock (2001; ZS2001) and the standard composition of seawater (Millero *et al.*, 2008; Millero2008). The Kargel2000 and ZS2001 K1 compositions are both enriched in Mg and $SO_4$, with similar relative proportions of Na, Ca, and Cl such that the resulting brine compositions are identical except for the presence of K in ZS2001. Sulfate fractionates out of the ice in all calculations, regardless of the composition of the ocean.

Because fluid brines are only stable in the portion of the ice that is above the eutectic temperature, salts transported toward the surface are likely to be in the solid phase and enriched in chlorides even if drawn from a sulfate-rich ocean. This fractionation at the ice-ocean interface is observed in multiyear sea ice on Earth (Gjessing *et al.*, 1993; Maus *et al.*, 2011), in which washout—melting and outflow from the ice into the ocean—creates a deficit of sulfate in the ice. Fractionation is also observed in ice of the "brine zone" of the McMurdo ice shelf, with the upper portion of the seawater-fed shelf containing less than 10% of the original $SO_4^{2-}$ content (Cragin *et al.*, 1986).

The above insights seem to suggest that Europa's ice is inefficient at transporting sulfate to the surface. It is worth noting here that a predicted dearth of oceanic sulfate at Europa's surface concurs with ground-based near-infrared observations with the Keck telescope. These recent observations indicate that MgSO$_4$ occurs mainly as an exogenic product of irradiated sulfur from Io (Brown and Hand, 2013) and that sulfates appear to be absent from many regions of Europa where they had previously been inferred from lower-resolution *Galileo* spacecraft spectra (Fischer *et al.,* 2016). Contrary to the interpretations offered from these recent observations, however, the present calculations suggest that evidence from Europa's surface is insufficient to conclude that its ocean, like Earth's, mainly contains Na and Cl.

In the preceding discussion we have focused on the composition and occurrence of ionic brines drawn from seawater composition. We leave estimates of the total available content and transport to future work. We conclude the discussion of the ice's composition by emphasizing that the influence of trapped materials on the ice's dynamics and thermal properties may be profound, even more so because the diversity of captured materials extends beyond the materials we have considered thus far. Among the many exotic brine compositions that may occur near Europa's surface, mobilized materials may draw from mixtures of exogenic sulfuric acid, perchlorates, ammonia, or methanol, which can have much lower melting temperatures than chlorides (Kargel*,* 1992; McCarthy *et al.,* 2007) and span many orders of magnitude in viscosity (Kargel *et al.,* 1991).

The growth and stability of brinicles in icy ocean worlds may be limited by turbulence at the ice-ocean interface. Whereas it has been suggested that freshwater dynamics may lead to a stratified layer at the interface, such as occurs in freshwater subglacial lakes on Earth (Melosh *et al.,* 2004), more recent investigations suggest that turbulent convection can dominate even for relatively low Rayleigh number flows, and these may disrupt stratified layers at the ice-ocean interface (Soderlund *et al.,* 2014; Jansen, 2016). These studies differ in their predictions. Most notably, the 100× slower flow speeds (~cm s$^{-1}$) predicted by Jansen (2016) may permit the formation of melt lenses under the ice (Zhu *et al.,* 2017) that provide stably stratified settings where brinicles can persist.

## 5. A Fluid Mechanical Model

Gravity on Europa is 1/7th $g_{Earth}$ ($g_{Europa}$ = 1.3 m/s² at the surface), and on Enceladus it is 1/90th $g_{Earth}$ ($g_{Enceladus}$ = 0.113 m/s²). Can we predict the relative lengths and thicknesses of brinicles on these bodies, neglecting the range of possible compositions and considering only the Earth-composition analog?

Let us begin by considering a minimal model of the radius of fluid-jet precipitated tubes. An approximate 1D analytical solution to the Navier-Stokes equations in the laminar flow regime leads to a parallel-velocity flow model for the radius and flow rate of a cylindrical jet of fluid that forms the template for the growth of a tube precipitated about itself (Cardoso and Cartwright, 2017)

$$\frac{8\mu_i}{\Delta\rho g\pi R_C^4}Q_i = R'^4\left[1 + 4\frac{\mu_i}{\mu_e}\left(\ln\frac{1}{R'} - \frac{1}{2}\right) + \frac{1}{\Delta\rho_i g}\frac{dP}{dz}\left(\frac{\mu_i}{\mu_e}\left(2 - \frac{1}{R'^2}\right) - 1\right)\right] \quad (1)$$

Here $R'$ is the radius of the tube growing in an environment of radius $R_C$, where this depends on the radius of external recirculation in the environment (see Cardoso and Cartwright, 2017), $\Delta\rho_i$ is the density difference between the brine and the external fluid, $\mu_i$ and $\mu_e$ are the internal and external fluid viscosities, and $dP/dz$ is the longitudinal pressure gradient. We have compared the prediction of this equation with the results of laboratory experiments on growing brinicles by Martin (1974) (Fig. 5). As seen, most points lie within experimental error along the Poiseuille flow solution, which is contained within Eq. 1 in the limit of large external fluid viscosity. This simple model provides a good starting point for understanding brinicle growth: for brinicles, the inner flow is affected only by the presence of the wall, so that effectively the outside fluid is solid and Poiseuille flow applies. This is in contradistinction to the growth of chemical gardens in the laboratory, where a similar comparison of Eq. 1 with data shows that the tube wall in that case behaves as a liquid, rather than a solid (Cardoso and Cartwright, 2017). Among other things, then, Eq. 1 tells us that the volumetric flow rate determines the diameter of the brinicle downflow tube.

Alongside this argument for the tube diameter, let us also consider the wall thickness. The thickness of the growing tube wall should scale with $\sqrt{(D\,t)}$, but in a chemical garden the diffusivity $D$ is a chemical diffusivity, while in a brinicle it is a thermal diffusivity, typically 100× larger. Therefore, a brinicle should increase in thickness faster than a laboratory chemical garden. We can check this prediction with chemical garden data from Stone *et al.* (2005): $t \sim 100$ min, thickness $\sim$1–2 mm; and brinicle data from Martin (1974): $t \sim 6$–10 min, thickness $\sim$5 mm. Hence, the radial diffusivity is, for chemical gardens, approximately $D \sim 0.001^2/(100 \cdot 60) = 1.7 \times 10^{-9}$ m$^2$/s and for brinicles $D \sim 0.005^2/(6 \cdot 60) = 7 \times 10^{-8}$ m$^2$/s. These data then confirm our scaling prediction. So brinicles grow by thermal diffusion, $\sim$100× faster than the chemical diffusion in chemical gardens. This contrasting growth speed underpins the difference in scales between chemical gardens and brinicles and the pure Poiseuille flow behavior we find in the brinicle compared to the "liquid-wall" behavior in the chemical gardens (Cardoso and Cartwright, 2017).

Let us now apply these arguments to brinicles on Europa and Enceladus. Firstly, the wall thickness scales as $\sqrt{(D\,t)}$, where $D$ is the thermal diffusivity; that is, this aspect is independent of gravity. Secondly, from Eq. 1, for a given brinicle radius, the flow rate will be lower in lower gravity, and consequently the brinicle will grow more slowly. Hence, a brinicle of a given length will be older in lower gravity, and its walls will be correspondingly thicker. We see from Eq. 1 that the precise way in which this occurs depends on the interplay between pressure-driven and buoyancy-driven flow in the brinicle; that is, to what extent brine flow is forced versus buoyancy-driven. Altering $g$ alters one part of the equation—the buoyancy-driven part—but does not affect pressure-driven flow.

The maximum effect of lower gravity comes when the pressure-driven component is negligible. In that case, we can affirm that for the same rate of flow, a brinicle would be $\sqrt[4]{7} \sim 1.6$ times larger on Europa and $\sqrt[4]{90}$ ~ 3 times larger on Enceladus (Fig. 5). This is likely so for brinicles developing beneath a growing floating ice sheet as depicted in Fig. 2, where brine accumulates at the freezing interface at the base of the ice. Thus, brine outflow at the base of the ice will occur at a rate determined by the buoyancy-driven flux at the outflow point. Compared with a similar concentration of outflowing brine on Earth, a brine outflow on Europa, with smaller *g*, will move more slowly. At the point that the outflow enters the ocean, a strong temperature gradient is established, but the difference between the outflowing fluid temperature and that of the surrounding ocean should diminish in the growing chimney with distance away from the outflow point, to the place where the down-flowing fluid melts chimney material at the same rate that the ice forms. Entrainment of freshwater at this point will mean that the outflow becomes less negatively buoyant and begins to spread laterally.

To consider the above flow rates in the context of possible fluid sources, we adopt the volume of a putative near-surface brine lens on Europa (Schmidt *et al.,* 2011), perched atop a layer of ductile convective ice. This potential reservoir of fluid is of critical astrobiological import as a potential means for delivering radiolytically produced oxidants and exogenously delivered nutrients to the ocean. While a direct connection to the ocean seems unlikely in this scenario, an indirect connection may be created by entrainment of briny fluids in pores and subsequent transport of those fluids downward along the near-melting thermal profile of the ice (Kalousová *et al.,* 2014). We consider the timescale for draining the feature,

$$\tau = V/Q \quad (2)$$

Supposing the lens to be a cylinder with a radius $R = 10$ km and depth $D = 100$ m, the entire volume of the chaos feature ($V = \pi R^2 D$) could be delivered to the ocean through a brinicle with $R \leq 25$ cm ($R_c = 10$ m) in less than 10 kyr. By contrast, prior studies suggest that the minimum transit time through the ice is greater than 10 kyr (Sotin *et al.,* 2002; Kalousová *et al.,* 2014). The insufficient supply of brines from the near surface has a few possible answers: brinicles undergo periods of inactivity; form, decay, and reform stochastically with the release of fluids; or a single brinicle serves as a debouché for multiple brine reservoirs. As discussed above, oceanic brines trapped within the lower 10% of the ice may also have a key role in brinicle formation, on Europa or on other ocean worlds such as Enceladus.

## 6. Conclusions

Brinicles provide a plausible setting for geochemical gradients amenable to life at the ice-ocean interface (Russell *et al.,* 2017), in some ways analogous to hydrothermal vents at the seafloor-ocean interface. We demonstrated here how brinicles may grow by thermal diffusion, and we provided a simple scaling for their growth and outflow rates in the absence of background turbulence. The broad range of possible compositional and dynamical configurations of the ice and ocean on Europa underscores uncertainties in predicting how and where brinicles may occur. As demonstrated by the above calculations and analogies to ices on Earth, fractionation of brines within Europa's ice may confound direct inference of the ocean's sulfate composition based solely on surface observations. Determining the possible existence of brines on other worlds requires understanding the detailed workings of their icy lithospheres: the generation and transport of fluids, and the role of salts. Promising measurements by NASA's planned Europa Clipper mission (Pappalardo *et al.,* 2016) and ESA's JUICE mission (Grasset *et al.,* 2013) that may contribute to this at Europa include the application of ice-penetrating radar; ; characterization of endogenous materials in the exosphere and potential plumes; investigations of ocean conductance using magnetometry; and detailed imaging to map the surface geology, composition, and thermal state. A seismic investigation on a follow-on lander, as recently studied by NASA (Hand *et al.,* 2017), could further constrain the properties of the ice and its interaction with the underlying ocean, and could potentially search for signatures of brinicles (Vance *et al.,*

2018).

## Acknowledgments

The authors thank Norm Sleep and two anonymous reviewers for helpful input, and Mohit Melwani Daswani for additional input on the nearly final manuscript. Brine calculations benefitted from work by JPL interns Nina Bothamy and Amira Elsenousy. S.D.V. thanks Bruce Bills and Baptiste Journaux for many stimulating discussions relevant to this work. Research by L.M.B. and S.D.V was carried out at the Jet Propulsion Laboratory, California Institute of Technology, under a contract with the National Aeronautics and Space Administration. L.M.B. and S.D.V. were supported by the NASA Astrobiology Institute (NAI) Icy Worlds project (13-NAI7-0024). S.S.S.C. acknowledges the financial support of the UK Leverhulme Trust project RPG-2015-002. J.H.E.C. acknowledges the Spanish MINECO grant FIS2016-77692-C2-2-P. No competing financial interests exist. © 2018. All rights reserved.

Address correspondence to:
*Steven D. Vance*
*Jet Propulsion Laboratory*
*California Institute of Technology*
*MS 321-560 4800 Oak Grove Drive*

*Pasadena, CA 91109*

*E-mail:* svance@jpl.nasa.gov

**Abbreviations Used**

| | |
|---:|---|
| Kargel2000 = | Kargel *et al.* (2000) |
| M2005 = | Marion *et al.* (2005) |
| Millero2008 = | Millero *et al.* (2008) |
| ZS2001 = | Zolotov and Shock (2001) |

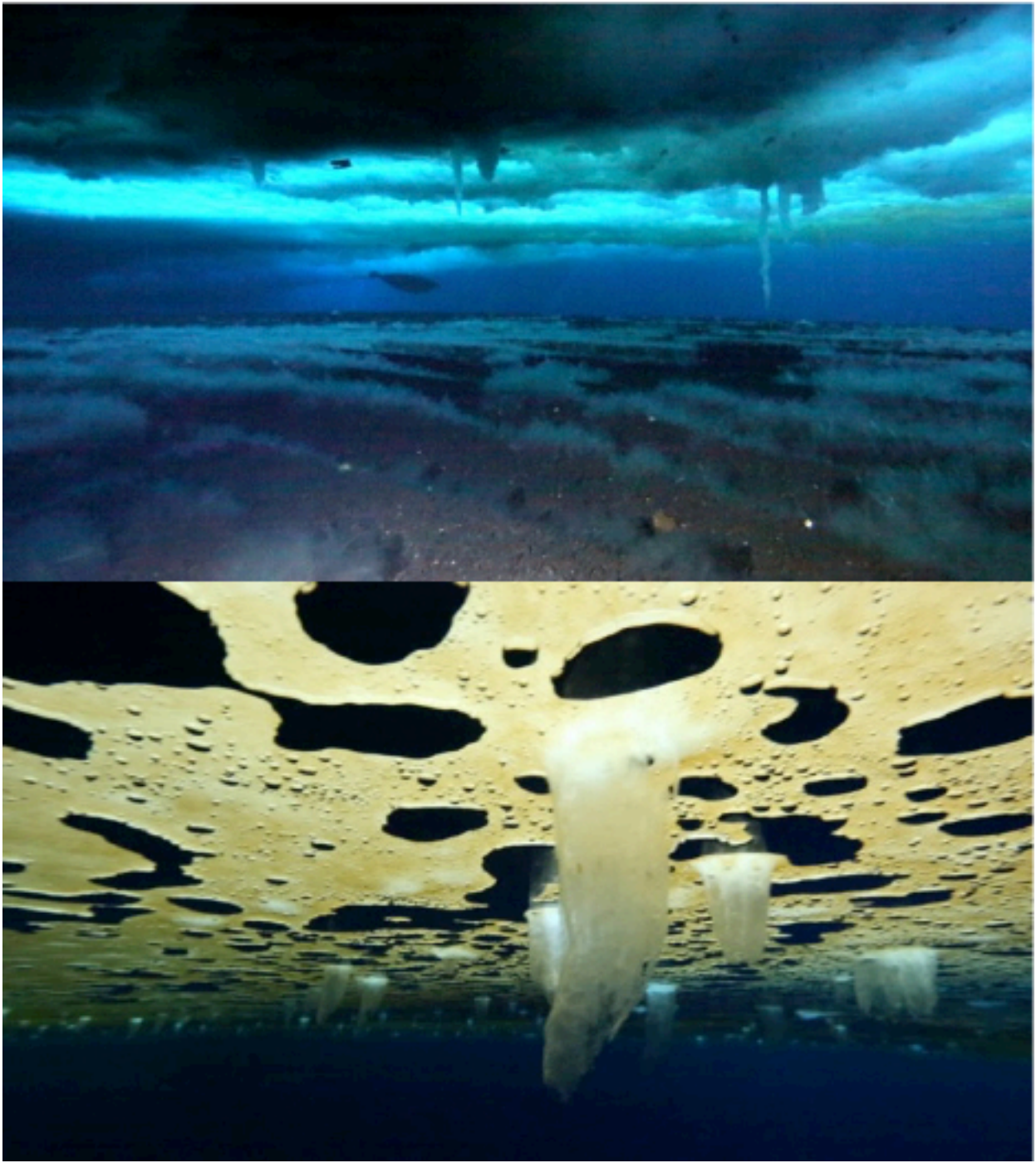

**FIG**. **1**. Brinicles under sea ice near McMurdo Station of the US Antarctic Survey. Images courtesy of Rob Robbins.

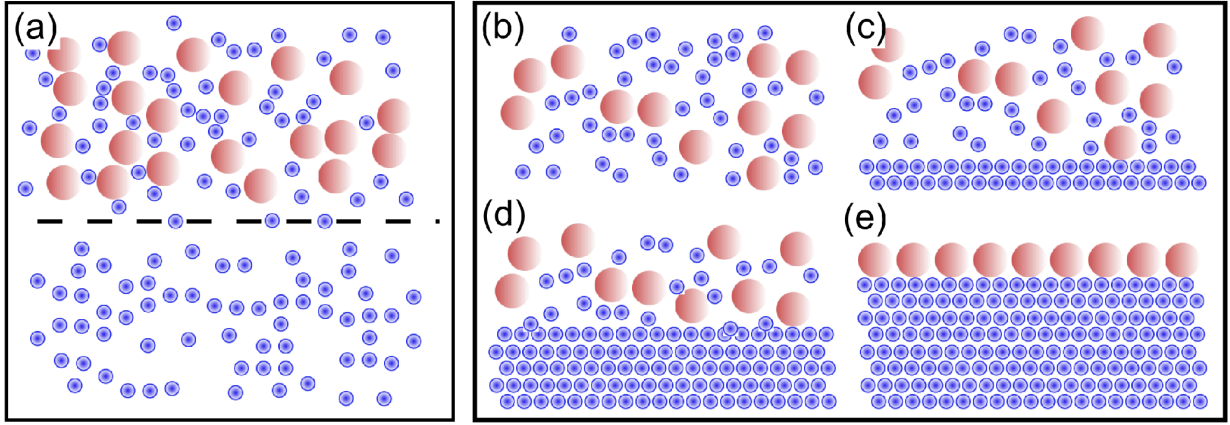

**FIG. 2**. Process of brinicle formation. The brine, concentrated by osmotic salt rejection as the ice sheet freezes (**a**), siphons successively (**b**–**d**) through a network of brine channels and injects downward into the surrounding ocean (**e**). From Cartwright *et al.* (2013).

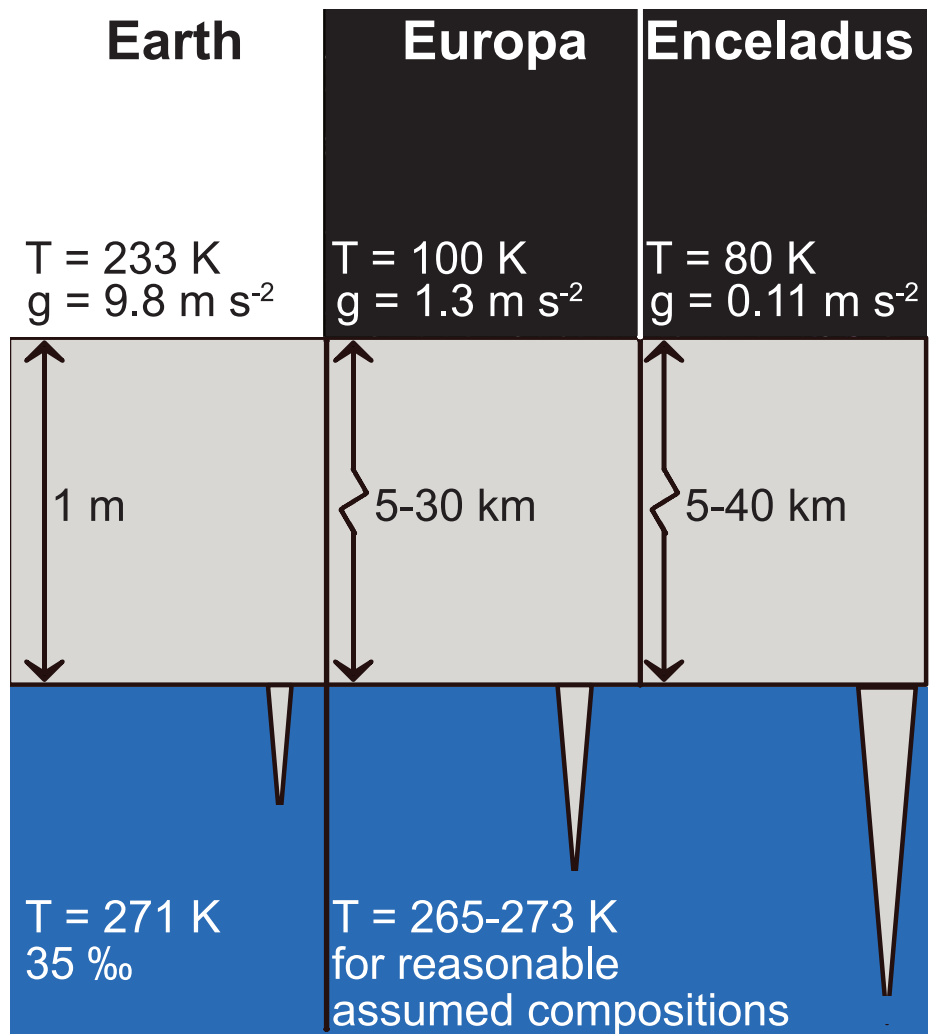

**FIG**. **3**. Seawater freezes on contact with this supercooled brine, forming a hollow tube. Comparison of surface, ice, and ocean conditions on Earth, Europa, and Enceladus, with relative scaling of the brinicles, and absolute scaling relative to the ice for Earth.

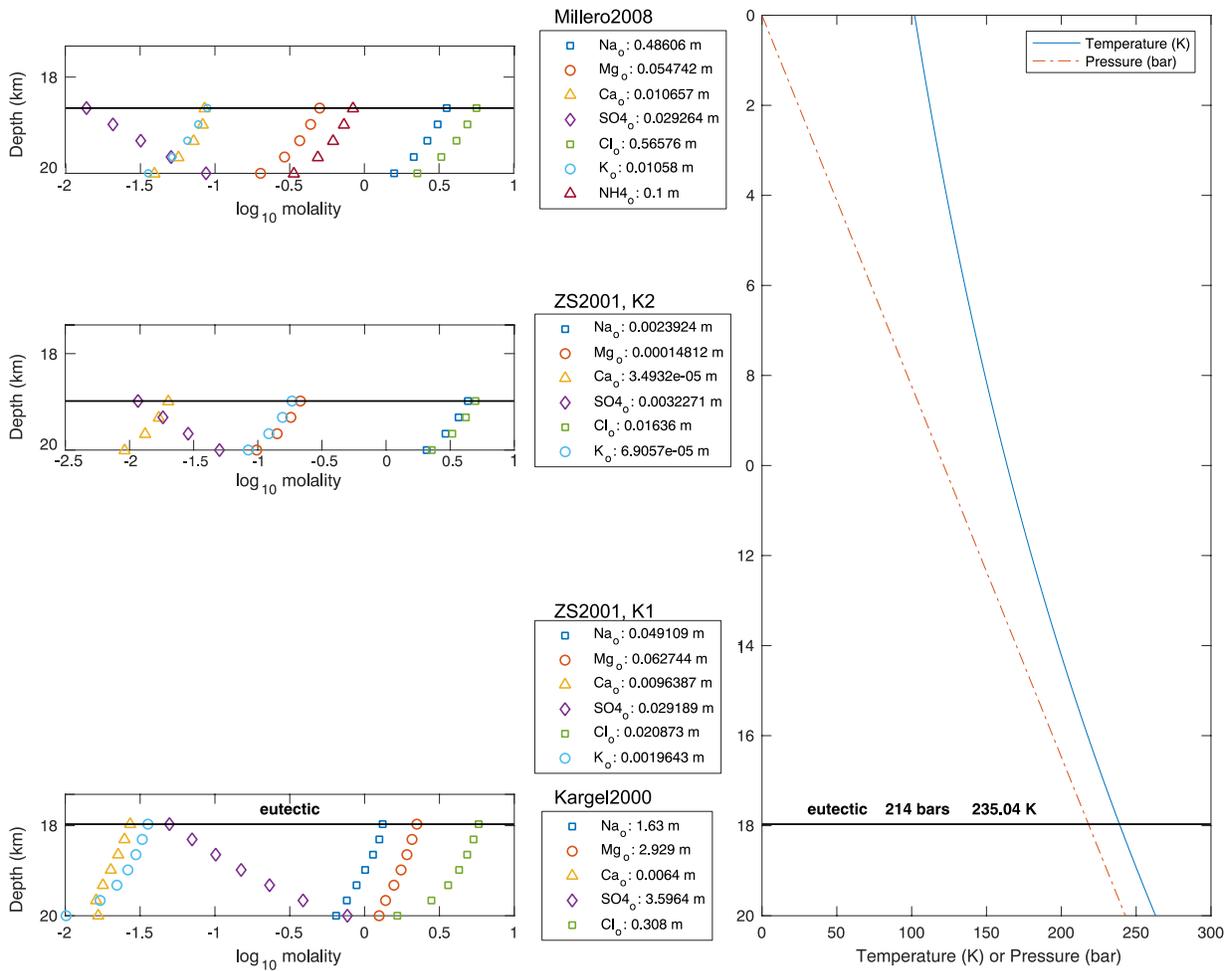

**FIG. 4**. Equilibrium brine content of a 20 km thick ice shell on Europa in the absence of solid-state convection, after Marion *et al.* (2005). Pressure and temperature in the thermally conductive ice are shown on the right (dashes and solid, respectively). Magnesium- and sulfate-dominated compositions (bottom left) from Kargel *et al.* (2000) and Zolotov and Shock (2001) produce identical equilibrium brine contents within the ice. For these, and for the sodium- and chlorine-dominated ocean compositions from Zolotov and Shock (2001; middle left) and the standard composition of seawater (Millero *et al.*, 2008; upper left), the overall effect of equilibrium freezing is to fractionate out $SO_4$.

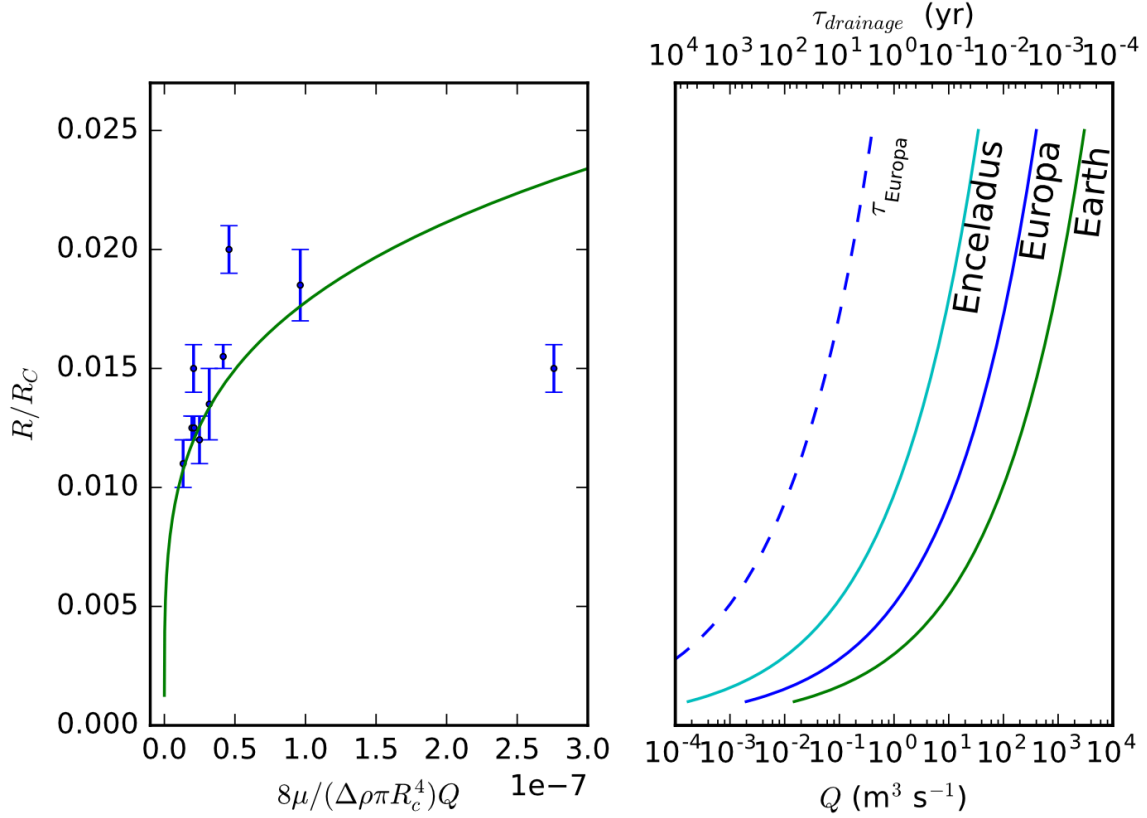

**FIG. 5**. Variation of tube radius ($y$ axis) with flow rate for brinicles (Eq. 1; Cardoso and Cartwright, 2017). **Left**: nondimensional flow relative to the circulation length scale $R_c$ (Eq. 1). Martin's (1974) experimental data for $R_c$ = 10 cm, $\Delta\rho_i$ = 150 kg m$^{-3}$, $\mu_i$ = 1.03 cP, $\mu_i/\mu_e$ = 1, $g$ = 9.81 m s$^{-2}$ are shown with the prediction from Poiseuille flow (Eq. 1) with $dP/dz = 0$ and $\mu_e \to \infty$.

**Right**: dimensionalized flow for Earth, Europa, and Enceladus ($R_c$ = 10 m, $\Delta\rho_i$ = 150 kg m$^{-3}$, $\mu_i$ = 1.03 cP, $\mu_i/\mu_e$ = 1, $g_{Earth}$ = 9.81 m s$^{-2}$, $g_{Europa}$ = 1.3 m s$^{-2}$; $g_{Enceladus}$ = 0.113 m s$^{-2}$). The dashed line depicts the drainage time for a hypothetical chaos lens on Europa modeled as a cylinder that is 100 m deep and 20 km in diameter ($\tau = V/Q$).